\documentclass[conference,10pt]{IEEEtran}
\usepackage{amsmath}
\usepackage{bm}
\usepackage{algorithmic}
\usepackage{hyperref}
\usepackage{algorithm}
\usepackage{subfigure}
\usepackage{graphicx,epsfig,balance}
\usepackage{cite}
\usepackage{color}
\usepackage{multirow}
\usepackage{amsfonts,amssymb}
\usepackage{booktabs}
\usepackage{stfloats}
\usepackage{graphicx}
\makeatletter
\renewcommand{\maketag@@@}[1]{\hbox{\m@th\normalsize\normalfont#1}}%
\makeatother

\begin{document}

\title{Analysis of the Power Imbalance in Power-Domain NOMA on Correlated Rayleigh Fading Channels}

\author{Shaokai Hu$^\dag$, Hao Huang$^\dag$, Guan Gui$^{\dag,*}$, and Hikmet Sari$^{\dag,\ddag,*}$ \\
\\
$^\dag$College of Telecommunications and Information Engineering, NJUPT, Nanjing, China\\
$^{\ddag}$Sequans Communications, 15--55 Boulevard Charles de Gaulle, 92700 Colombes, France\\
$^{*}$E-mails: guiguan@njupt.edu.cn, hikmet.sari@centralesupelec.fr}

\maketitle
\begin{abstract}
This paper analyzes the power imbalance issue in power-domain NOMA (PD-NOMA) in the presence of channel correlations, typically encountered on the downlink of cellular systems when the base station antennas have an insufficient separation. In a recent paper, the authors analyzed this issue for a typical uplink scenario with uncorrelated channels, and the study revealed an astounding result that the optimum in terms of average error probability is achieved when the user signals are perfectly balanced in terms of power as in multi-user MIMO with power control. This result led to some questioning of the concept of PD-NOMA for uncorrelated Rayleigh fading channels. In the present paper, we make a similar analysis for the downlink, and the study gives a very clear insight into the influence of the power imbalance at different levels of channel correlation. First, with full correlation (user signals transmitted from the same antenna), the PD-NOMA concept reduces to simple signal constellation design. The optimum is achieved when the power imbalance between the user signals is such that the resulting constellation has uniform spacing. Any deviation from this optimum will lead to a hierarchical constellation with performance loss. Also, this optimum power imbalance is shown to hold for a range of strong channel correlations, but for moderate and low correlation values perfectly power balanced NOMA takes over as in the presence of uncorrelated channels.
\end{abstract}


\section{Introduction}
Non-orthogonal multiple access (NOMA) is currently viewed as one of the key physical layer (PHY) technologies for next generations of wireless networks. Since 2013, there has been a huge amount of literature on the subject, and the vast majority of this literature has been devoted to the so-called power-domain NOMA (PD-NOMA), although there exists other NOMA categories such as code-domain NOMA, interleave-domain NOMA, and other. PD-NOMA is based on imposing a power imbalance between user signals and detecting these signals using a successive interference cancellation (SIC) receiver (see, e.g., \cite{YS,ZD,LiuIEEE2017,LD,ZDING,XLEI,SY,MS,GuiCL2019a}). For a comprehensive review and description of other NOMA categories, the reader can refer to \cite{MV}. Another NOMA concept that is worth to mention is NOMA-2000 \cite{HSA,AME,AAX,IC}, which is based on using two sets of orthogonal signal waveforms \cite{HSLETTER,HSMAGAZINE} and iterative interference cancellation. This technique does not require any power imbalance between user signals and falls in the category of code-domain NOMA due to the fact that the signal waveforms in one of the two sets are spread in time or in frequency.

Well before the surge of literature on NOMA during the past decade, the concept of multiple-input multiple-output (MIMO) was generalized to multi-user MIMO (MU-MIMO), where the multiple antennas are not employed by the same user, but instead they correspond to multiple users \cite{QH,AM,SK,XC,BF}. In a cellular system, two users each of which is equipped with a single antenna and simultaneously communicating with a base station (BS) equipped with multiple antennas form a MU-MIMO system. Note that MU-MIMO is also known as virtual MIMO as it can be seen from the titles of \cite{SK,XC,BF}. Naturally, the question arises as to what relationship PD-NOMA has with the MU-MIMO concept and how it compares to it in terms of performance. In both techniques, user signals are transmitted in parallel using the same time and frequency resources and the only difference is that these signals have different powers in PD-NOMA while no power difference is involved in MU-MIMO. Therefore, MU-MIMO is a particular NOMA scheme which can be referred to as power-balanced NOMA or equal-power NOMA.

In a recent paper \cite{HSK}, focusing on a 2-user uplink with uncorrelated Rayleigh fading channels, the present authors gave a unified presentation of PD-NOMA and MU-MIMO by introducing a power imbalance parameter in the system model and optimized this parameter to minimize the average bit error probability for a given total transmit power by the two users. The study revealed a most striking result. Specifically, it was found that the optimum in this scenario corresponds to zero power imbalance, i.e., to the transmission of equal average power by the users and equal power received by the BS from each of them. This means that the power imbalance, which is the basic principle of PD-NOMA, is actually undesirable when the channels are uncorrelated, and PD-NOMA appears to be an ill concept in this scenario.

In the present paper, we investigate the power imbalance issue when the channels are correlated, which is typically the case on a cellular downlink with a limited separation between BS antennas. We consider different correlation levels going from full correlation (which corresponds to the transmission of user signals from the same antenna) to zero correlation, which was the assumption made in \cite{HSK}. Between these two extremes, the study covers strong correlation, medium correlation, and low correlation, and it gives significant insight into the influence of the power imbalance at these correlation levels. Our analysis leads to several important findings. First, the full correlation case corresponds to pure constellation design. In other words, the concept of PD-NOMA is completely reduced to designing a signal constellation with uniform spacing in order to maximize the minimum Euclidean distance and bit error rate (BER) performance for a given transmit power. Any deviation from the power imbalance leading to a uniform spacing in the resulting constellation will lead to a hierarchical constellation with a reduced minimum distance. Note that hierarchical constellations were introduced in the past for unequal error protection (see, e.g., \cite{MJ,AMC}), i.e., to have an increased error protection of some information bits with respect to others. Our second finding is that the optimum power imbalance parameter determined in the full correlation case also leads to the best performance results on channels with strong correlation. As the correlation parameter is reduced, the situation becomes very close to the zero correlation case, where the best strategy consists of perfect power balancing. We illustrate this behavior using the QPSK and 16QAM signal formats. A third interesting finding of this study is that the influence of channel correlation is highest in the power balanced scenario, and it virtually disappears when the power imbalance between user signals is very strong.

The paper is organized as follows. In Section \ref{sec2}, we briefly recall the principle of PD-NOMA and MU-MIMO and give a unified system model which includes channel correlation. Next, in Section \ref{sec3}, we focus on the full correlation case and we give the optimum power imbalance factor for the QPSK and 16QAM signal formats. We also evaluate the performance degradation caused by any deviation from these optimum values. In Section \ref{sec4}, we report the results of computer simulations showing BER performance for different channel correlation levels and power imbalance factors. Finally, Section \ref{sec5} gives our conclusions.

\section{System Model}
\label{sec2}
\subsection{Power-Domain NOMA}
\begin{figure}[htbp]
  \centering
  \includegraphics[width=3.3 in]{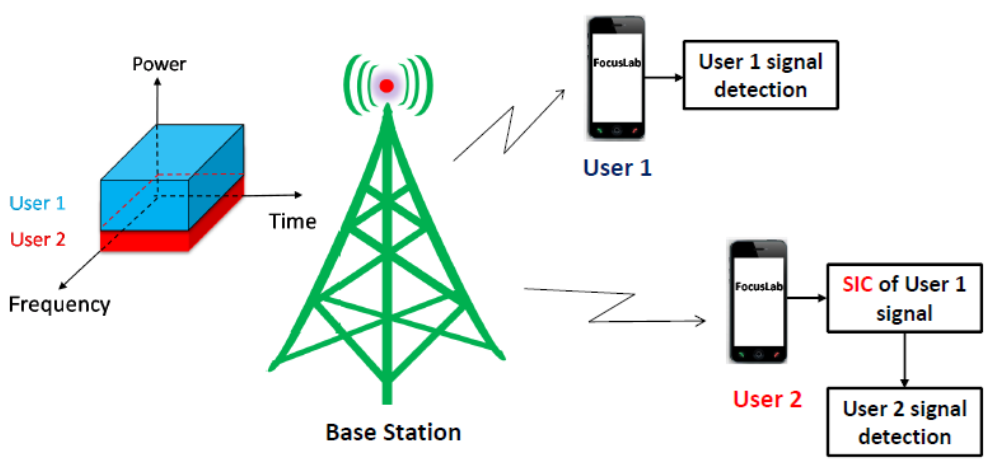}
  \caption{Illustration of PD-NOMA downlink with 2 users.}
  \label{fig1}
\end{figure}

The principle of PD-NOMA is to transmit user signals in the same time and frequency resource blocks by allocating different powers to them. On the receiver side, a SIC receiver is employed to detect the user signals \cite{DT}. Fig. \ref{fig1} shows the concept of a 2-user PD-NOMA downlink. The BS transmits a high signal power (shown in blue) to User 1 and a low signal power (shown in red) to User 2. Assuming that the power imbalance between the blue and the red signals is sufficiently high, User 1 can detect its signal in the presence of interference from the User 2 signal without a strong BER degradation. In contrast, User 2 cannot detect its signal directly. It must first detect the User 1 signal and subtract the interference of this signal before detecting its own signal. It is clear that detection of User 1 signal leads to a poor BER performance if the power imbalance between the two signals is not sufficiently high, and the SIC receiver fails to properly cancel the interference leading to degraded performance also for User 2 in this case. For this reason, user pairing in PD-NOMA is an important function, which aims at reducing performance degradation with respect to interference-free transmission. While the use of a SIC receiver is a common practice in the NOMA literature, some authors recently studied the use of maximum-likelihood (ML) detection and the performance improvement that can be achieved using this technique \cite{JS,HS}.

\subsection{Multi-User MIMO}
Multi-user MIMO is the terminology given to a MIMO system when the multiple antennas do not correspond to a single user. For example, a cellular system in which a BS equipped with multiple antennas communicating with a number of single-antenna users forms a MU-MIMO system. Note that with respect to conventional point-to-point MIMO links, MU-MIMO is what orthogonal frequency-division multiple access (OFDMA) is to orthogonal frequency-division multiplexing (OFDM). The first is a point-to-point transmission technique, and the second is a multiple access technique based on the same principle. As in the PD-NOMA briefly outlined in the previous subsection, here too we will focus on the downlink of a 2-user MU-MIMO system and assume that signals are transmitted to the two users using the same time and frequency resources simultaneously. In fact, Fig. \ref{fig1} can also be used to describe this MU-MIMO system if the user signals have equal power and the BS is equipped with multiple antennas, which is standard in state-of-the-art cellular networks. Therefore, MU-MIMO can be viewed as an equal-power NOMA system, in which the SIC receiver is not appropriate for signal detection. For both PD-NOMA and MU-MIMO, the optimum receiver is in fact the ML receiver, which makes its decisions by minimizing the Euclidean distance from the received noisy signal. In PD-NOMA with a large power imbalance between user signals, the SIC receiver essentially provides the ML detector performance, but this concept is not applicable to a power-balanced system.

\subsection{Unified System Model}
We now give a unified simple model for the downlink in 2-user PD-NOMA and MU-MIMO systems with one or two antennas on the user side and two antennas on the BS side. We consider the transmission of a symbol $x_1$ by the first BS antenna to the first user and the transmission in parallel of a symbol $x_2$ from the second BS antenna to the second user. Both symbols are assumed to take their values from the same signal constellation. A power imbalance factor $\alpha$ such that $1/2\le\alpha<1$ will be introduced with the value $\alpha=1/2$  corresponding to MU-MIMO and $\alpha\neq1/2$ corresponding to PD-NOMA. Note that if the BS is equipped with one antenna only, both symbols are transmitted from the same antenna, and the MU-MIMO system becomes a multi-user single-input multiple-output (MU-SIMO) system. The users are assumed to use ML detection to estimate the symbols intended to them, and assuming that the channels have the same statistics, it is sufficient to examine the receiver of only one of the users. If the device of this user is only equipped with one antenna, the received signal can be written as:
\begin{align}
\label{eq1}
r_{1}=\sqrt{\alpha }h_{11}x_{1}+\sqrt{1-\alpha }h_{12}x_{2}+w_{1}
\end{align}
where $h_{11}$ and $h_{12}$ denote the responses of the channels between the BS antennas and the first antenna of this user, and $w_1$ is an additive white Gaussian noise (AWGN) term. If the user device is equipped with two antennas, the signal received by the second antenna can be similarly written as:
\begin{align}
\label{eq2}
r_{2}=\sqrt{\alpha }h_{21}x_{1}+\sqrt{1-\alpha }h_{22}x_{2}+w_{2}
\end{align}
where $h_{21}$ and $h_{22}$ denote the responses of the channels between the BS antennas and the second antenna of this user, and $w_2$ is an AWGN term.

The channels are assumed to be unity-variance Rayleigh fading channels, and under this assumption the power imbalance at the receiver is identical to that imposed at the BS side. In the case the BS is equipped with a single antenna and both symbols are transmitted from this antenna, the users' signals propagate on the same channels and by definition the power imbalance is the same on both sides.
Let us combine equations (\ref{eq1}) and (\ref{eq2}) and write the vector equation:
\begin{align}
\label{eq3}
R=HX+W
\end{align}
where $R=\begin{pmatrix}
r_{1} \\r_{2}
\end{pmatrix}, H=\begin{pmatrix}
 h_{11}&h_{12}  \\
 h_{21}&h_{22}  \\
\end{pmatrix}, X=\begin{pmatrix}
\sqrt{\alpha }x_{1} \\\sqrt{1-\alpha }x_{2}
\end{pmatrix}$, and $W=\begin{pmatrix}
w_{1} \\w_{2}
\end{pmatrix}$.

In this description, the total transmit power is $\sigma _{x}^{2}=E\left ( \left|x_{1} \right|^{2} \right )=E\left ( \left|x_{2} \right|^{2} \right )$, a fraction of $\alpha\sigma _{x}^{2}$ being associated to the User 1 signal and a fraction of $(1-\alpha)\sigma _{x}^{2}$ being associated to the User 2 signal. The ML receiver makes its decisions by minimizing the Euclidean distance of the received noisy signal $R$ from $HX$ over all values of the symbol vector $X$. This can be written as:
\begin{align}
\label{eq4}
\hat{X}=\arg\displaystyle\min_X\begin{Bmatrix}
\left\|R-HX \right\|^{2}\end{Bmatrix}
\end{align}
For a constellation size $M$, ML detection involves the computation of  $M^2$ metrics and their comparisons in order to find the minimum value. Note that the complexity is significantly higher than that of the SIC receiver, which only involves the computation and comparison of $2M$ metrics \cite{HS}.

\section{Power Imbalance and Channel Correlation}\label{sec3}
We will now analyze the power imbalance issue of PD-NOMA in the presence of channel correlation. Correlation will refer to correlation between transmit antennas, and we will assume that there is no correlation between the receive antennas. Referring back to equations (\ref{eq1}) and (\ref{eq2}), the correlation is between $h_{11}$ and $h_{12}$, and also between $h_{21}$ and $h_{22}$, but there is no correlation between the first pair and the second pair of channel responses. Accordingly, we generate 4 uncorrelated Rayleigh fading channel coefficients $f_1, f_2, f_3, f_4$, and from these we generate the correlated channel coefficients as:
\begin{align}
\label{eq5}
h_{11}=\frac{1}{\sqrt{1+\gamma ^{2}}}\left ( f_{1}+\gamma f_{2} \right ), h_{12}=\frac{1}{\sqrt{1+\gamma ^{2}}}\left ( \gamma f_{1}+f_{2} \right )
\end{align}
and
\begin{align}
\label{eq6}
h_{21}=\frac{1}{\sqrt{1+\gamma ^{2}}}\left ( f_{3}+\gamma f_{4} \right ), h_{22}=\frac{1}{\sqrt{1+\gamma ^{2}}}\left ( \gamma f_{3}+f_{4} \right )
\end{align}
The correlation parameter $\gamma$ in these equations takes values from 0 to 1, with $\gamma=0$ corresponding to no correlation and $\gamma=1$ corresponding to full correlation.

We will start our analysis with full correlation, which is equivalent to transmission of the users' signals from the same antenna. In this case, we have $h_{12}=h_{11}$ and $h_{22}=h_{21}$, and equations (\ref{eq1}) and (\ref{eq2}) become:
\begin{align}
\label{eq7}
r_{1}=\left ( \sqrt{\alpha }x_{1} +\sqrt{1-\alpha }x_{2}\right )h_{11}+w_{1}
\end{align}
and
\begin{align}
\label{eq8}
r_{2}=\left ( \sqrt{\alpha }x_{1} +\sqrt{1-\alpha }x_{2}\right )h_{21}+w_{2}
\end{align}

As mentioned previously, it is assumed that the symbols $x_1$ and $x_2$ take their values from the same signal constellation, and we will focus on the QPSK and 16QAM constellations in this paper. Let us first point out that in the full correlation case, the use of equal power for the two symbols (i.e., $\alpha=1/2$) leads to singularities, and it is not possible to uniquely determine the transmitted symbols. To view this, consider a symbol $x_{1}^{'}=x_{1}+\delta _{0}$ transmitted to User 1 and a symbol $x_{2}^{'}=x_{2}-\delta _{0}$ transmitted by User 2, where $\delta _{0}$ denotes the minimum distance of the signal constellation. For $\alpha=1/2$, we have $\sqrt{\alpha }x_{1}^{'}+\sqrt{1-\alpha }x_{2}^{'}=\sqrt{\alpha }x_{1}+\sqrt{1-\alpha }x_{2}$ and it is impossible to uniquely determine the transmitted symbols upon reception of $r_1$ and $r_2$. For this reason, a power imbalance is needed in the full correlation case, and the optimum power imbalance in terms of average error probability is the one which leads to a combined constellation with a uniform spacing. In other words, the PD-NOMA concept reduces to a problem of constellation design in the full correlation case.

Let us determine the optimum power imbalance for the QPSK and 16QAM signal constellations. First, with the QPSK signal constellation in which the in-phase (I) and quadrature (Q) components take their values from the set $\begin{Bmatrix}\pm1\end{Bmatrix}$, the minimum distance of the constellation from which the transmitted signal $\sqrt{\alpha }x_{1}+\sqrt{1-\alpha }x_{2}$ takes its values is given by:
\begin{align}
\label{eq9}
d_{\min}=\min\left ( 2\sqrt{1-\alpha },2\left (\sqrt{\alpha }-\sqrt{1-\alpha }\right ) \right )
\end{align}

The distance is maximum when $\sqrt{\alpha }=2\sqrt{1-\alpha }$, i.e., when $\alpha=0.8$. This corresponds to a uniform 16QAM signal constellation which provides the same BER performance for the $x_1$ and $x_2$ symbols. If $\alpha\neq0.8$, the constellation is hierarchical, and the symbols $x_1$ and $x_2$ have unequal BER performances, with the $x_1$ symbol having better performance for $\alpha>0.8$ and the $x_2$ symbol having better performance for $\alpha<0.8$.
Next, with the 16QAM signal constellation in which the I and Q components take their values from the set $\begin{Bmatrix}\pm1,\pm3\end{Bmatrix}$, the minimum distance of the $\sqrt{\alpha }x_{1}+\sqrt{1-\alpha }x_{2}$ constellation is given by:
\begin{align}
\label{eq10}
d_{\min}=\min\left (2\sqrt{1-\alpha },2\left (\sqrt{\alpha }-3\sqrt{1-\alpha }\right ) \right )
\end{align}
The distance is maximum when $\sqrt{\alpha}=4\sqrt{1-\alpha }$, i.e., when $\alpha=16/17$, this value leading to a uniform signal constellation which provides the same BER performance for both users. As previously, better BER performance is achieved for User 1 symbols with $\alpha>16/17$, and better performance is achieved for User 2 symbols with $\alpha<16/17$.

\section{Simulation Results}
\label{sec4}

Using the QPSK and 16QAM signal constellations, a simulation study was performed to evaluate the influence of the power imbalance factor on the bit error rate (BER) of PD-NOMA on correlated Rayleigh fading channels. Following the system model of Subsection II.C and the channel correlation model of Section \ref{sec3}, we considered the downlink of a PD-NOMA system with two users. The receiver in the user equipment employs ML detection and assumes that the channel state information (CSI) is perfectly known. The simulations were performed using various correlation parameters ranging from full correlation $(\gamma=1)$ to no correlation $(\gamma=0)$, but only those corresponding to $3$ correlation parameters will be shown here due to space limitations. These correspond to full correlation $(\gamma=1)$, strong correlation $(\gamma=0.9)$, and mild correlation $(\gamma=0.5)$. Note that the results corresponding to no correlation $(\gamma=0)$ were previously reported in \cite{HSK}.

\begin{figure}[htb]
  \centering
  \includegraphics[width=3.4 in]{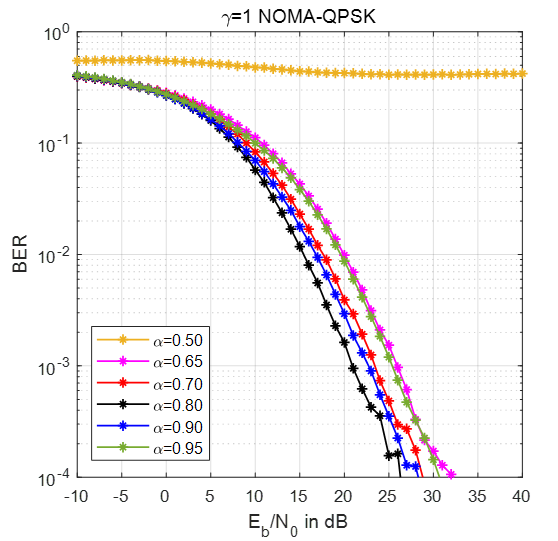}
  \caption{BER Performance of PD-NOMA with QPSK on fully correlated channels $(\gamma=1)$ and different values of the power imbalance factor $\alpha$.}
  \label{fig2}
\end{figure}
\begin{figure}[htb]
  \centering
  \includegraphics[width=3.4 in]{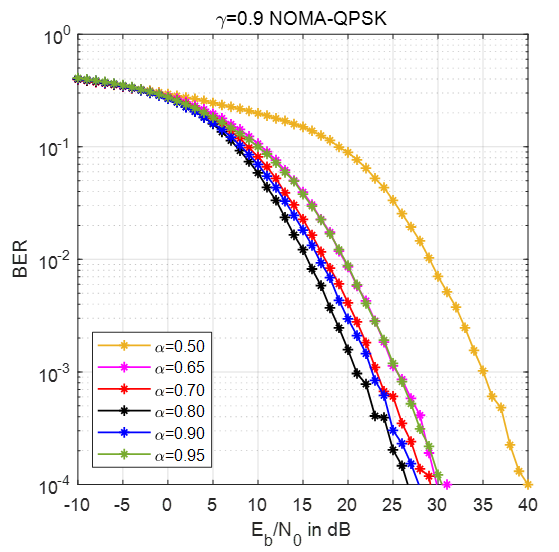}
  \caption{BER Performance of PD-NOMA with QPSK on strongly correlated channels $(\gamma=0.9)$ and different values of the power imbalance factor $\alpha$.}
  \label{fig3}
\end{figure}
\begin{figure}[htb]
  \centering
  \includegraphics[width=3.4 in]{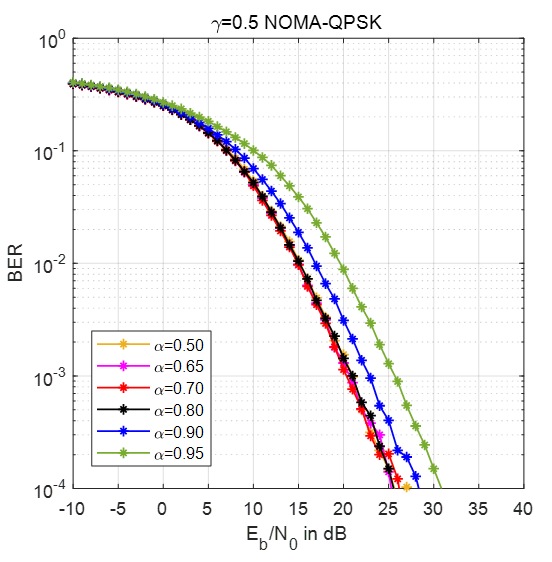}
  \caption{BER Performance of PD-NOMA with QPSK on mildly correlated channels $(\gamma=0.5)$ and different values of the power imbalance factor $\alpha$.}
  \label{fig4}
\end{figure}

The simulation results are reported in Fig. \ref{fig2} for QPSK on fully correlated channels. The figure shows that as predicted the best results correspond to power imbalance factor $\alpha=0.8$. It also shows an error floor of BER$=0.4$ for power imbalance factor $\alpha=0.5$. Finally, the figure also shows a signal-to-noise ratio (SNR) degradation on the order of $2\sim3$ dB for power imbalance factors of $\alpha=0.9$ and $\alpha=0.7$, and an SNR degradation on the order of $5$ dB for power imbalance factors of $\alpha=0.95$ and $\alpha=0.65$. Next, Fig. \ref{fig3} shows the results for QPSK on strongly correlated channels $(\gamma=0.9)$. Here too, we can see that the best results are obtained with the power imbalance factor $\alpha=0.8$ and that the power imbalance factors $\alpha=0.65$, $\alpha=0.7$, $\alpha=0.9$, and $\alpha=0.95$ lead to the same SNR degradations as in the full correlation case. Finally, the power imbalance factor $\alpha=0.5$ no longer leads to a BER floor, but the SNR degradation is still very large compared to the case with optimum power imbalance. The simulation results with QPSK on mildly correlated channels $(\gamma=0.5)$ are reported in Fig. \ref{fig4}. We can see that power imbalance factor $\alpha=0.5$ gives here the same results as the power imbalance factors $\alpha=0.65$, $\alpha=0.7$, and $\alpha=0.8$, while the power imbalance factor $\alpha=0.9$ leads to some $3$ dB SNR degradation, and the power imbalance factor $\alpha=0.95$ leads to some 5 dB SNR degradation. These results are quite close to those corresponding to $\gamma=0$ reported in \cite{HSK}, the basic difference being that the optimality of the power imbalance factor $\alpha=0.5$ appears very clearly in the case of no channel correlation.

\begin{figure}[htb]
  \centering
  \includegraphics[width=3.4 in]{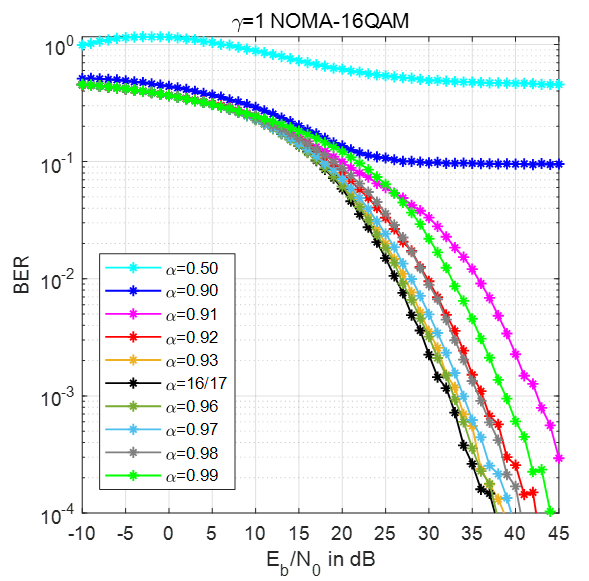}
  \caption{BER Performance of PD-NOMA with 16QAM on fully correlated channels $(\gamma=1)$ and different values of the power imbalance factor $\alpha$.}
  \label{fig5}
\end{figure}

\begin{figure}[htb]
  \centering
  \includegraphics[width=3.4 in]{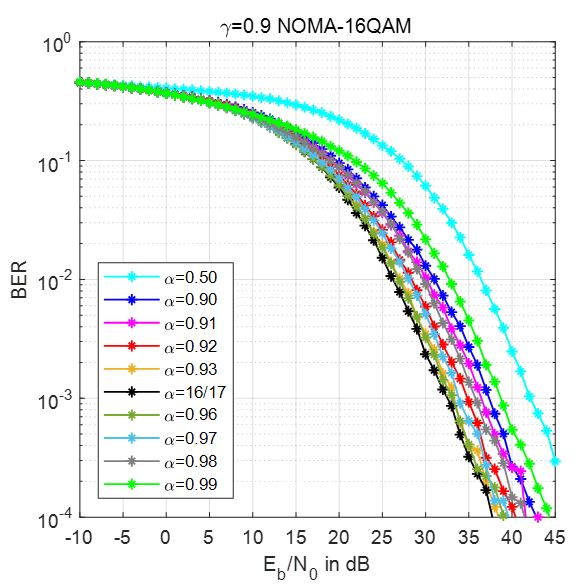}
  \caption{BER Performance of PD-NOMA with 16QAM on strongly correlated channels $(\gamma=0.9)$ and different values of the power imbalance factor $\alpha$.}
  \label{fig6}
\end{figure}
\begin{figure}[htb]
  \centering
  \includegraphics[width=3.4 in]{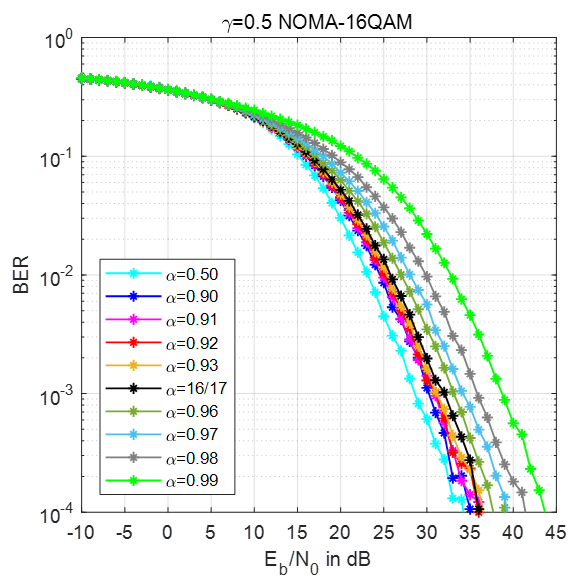}
  \caption{BER Performance of PD-NOMA with 16QAM on mildly correlated channels $(\gamma=0.5)$ and different values of the power imbalance factor $\alpha$.}
  \label{fig7}
\end{figure}

The simulation results with 16QAM on fully correlated channels are reported in Fig. \ref{fig5}. As predicted by the analytical results of Section \ref{sec3}, it shows that the best results are indeed obtained with the power imbalance factor $\alpha=16/17$, which leads to a signal constellation with uniform signal spacing. It also shows that the SNR degradation increases as the power imbalance factor deviation from this optimum value is increased. Note that in addition to the BER floor of approximately $0.5$ obtained with $\alpha=0.5$, we can also see a BER error floor of $10^{-1}$ corresponding to $\alpha=0.9$. Next, Fig. \ref{fig6} shows the results with a strong correlation coefficient of $\gamma=0.9$. Here too, the power imbalance factor $\alpha=16/17$ is optimum, and deviation from this value leads to significant SNR degradation. For example, the power imbalance factor $\alpha=0.9$ leads to an SNR degradation of approximately $5$ dB, and the the imbalance factor $\alpha=0.99$ leads to an SNR degradation as high as $6$ dB. Finally, the results in the presence of a mild correlation $(\gamma=0.5)$ are depicted in Fig. \ref{fig7}. They show that as in the absence of correlation, the best results are obtained with a power imbalance factor of $\alpha=0.5$. Compared to this, the power imbalance factor $\alpha=0.9$ leads to an SNR degradation of approximately $3$ dB, and the power imbalance factor $\alpha=0.99$ leads to a degradation higher than $10$ dB.

\section{Conclusion}\label{sec5}
In this paper, we analyzed the power imbalance between user signals on the downlink of a 2-user Power-Domain NOMA system in the presence of correlated Rayleigh fading channels. We highlighted the fact that when the user signals are fully correlated, e.g., when they are transmitted from the same BS antenna, the concept of PD-NOMA reduces to signal constellation design, and the optimum power imbalance corresponds to designing a constellation with uniform spacing. Deviations from this value of the power imbalance factor leads to a significant SNR degradation. We also found that the optimum power imbalance corresponding to full correlation is also valid for the cases with strong correlation. But in the case of moderate or mild correlation, the best results are obtained with perfectly power-balanced user signals as in the case of uncorrelated channels. These findings, together with those reported in an earlier paper for the uplink, shed light and give a clear insight on the potential and limitations of this multiple access technique, which is advocated for future generations of wireless communications networks.

\end{document}